\documentstyle{article}
\begin{document}
 \title{Consistent use of paradoxes in deriving constraints on the
dynamics of physical systems and of no-go-theorems}
\author{K. Svozil\\
 {\small Institut f\"ur Theoretische Physik}  \\
  {\small Technische Universit\"at Wien   }     \\
  {\small Wiedner Hauptstra\ss e 8-10/136}    \\
  {\small A-1040 Vienna, Austria   }            \\
  {\small e-mail: svozil@tph.tuwien.ac.at}}
\maketitle

\begin{abstract}

The classical methods used by recursion theory and formal logic to block
paradoxes do not work in quantum information theory.
Since quantum information can exist as a coherent superposition of the
classical
``yes'' and ``no'' states,
certain tasks which are not conceivable in
the classical setting can be performed in the quantum setting.
Classical logical inconsistencies do not arise, since
there exist fixed point states of the diagonalization operator.
In particular, closed timelike curves need not be eliminated in the
quantum setting, since they would not lead to any paradoxical outcome
controllability.
Quantum information theory can also be subjected to the treatment of
inconsistent information in databases and expert systems.
It is suggested that any two pieces of contradicting information are
stored and processed as coherent superposition. In order to be
tractable, this strategy requires quantum computation.

\end{abstract}

\begin{flushright}
{\scriptsize paradox.tex}
\end{flushright}

\noindent

This letter introduces two novel features of quantum information
theory. Physically, it is shown how quantum information
allows the consistent implementation of nonlocal correlations.
 Technically, a diagonalization operator
 is used to compute consistent fixed
point solutions to classical ``paradoxical'' tasks.
The implications for quantum recursion theory \cite{qrt} and
algorithmic information theory \cite{omega} as well as for database
applications will only be shortly sketched.

Classical information theory (e.g., \cite{hamming}) is based on the bit
as
fundamental atom. This classical bit, henceforth called
{\em cbit,} is in one of two
classical states.
It is customary to use the symbols ``$0$'' and ``$1$'' as the names of
these states. The corresponding classical bit states are denoted by the
symbols $0$ and $1$.

In quantum information theory (cf.
\cite{a:8,deutsch-85,f-85,peres-85,b-86,m-86,deutsch:89,deutsch:92}),
the most elementary unit of information,
henceforth called {\em qbit},
may be physically represented by a coherent
superposition
of the two states $\vert 0\rangle $ and $\vert 1\rangle$, which
correspond to the symbols
$0$ and
$1$, respectively.
The quantum bit states
\begin{equation}
\vert a,b\rangle =a\vert 0\rangle +b\vert 1 \rangle
\end{equation}
form a continuum, with
$ \vert a\vert^2+\vert b\vert^2=1$, $a,b\in { C}$.

\begin{figure}
\begin{center}
\unitlength=1.00mm
\special{em:linewidth 0.4pt}
\linethickness{0.4pt}
\begin{picture}(91.00,85.00)
\put(74.00,20.00){\line(-1,1){59.00}}
\put(76.00,17.00){\makebox(0,0)[cc]{$t_S$}}
\put(45.00,40.00){\framebox(3.00,3.00)[cc]{$1$}}
\put(80.00,21.00){\framebox(3.00,3.00)[cc]{$2$}}
\put(15.00,85.00){\line(0,-1){15.00}}
\put(15.00,70.00){\circle*{2.83}}
\put(6.00,5.00){\vector(1,0){8.00}}
\put(6.00,5.00){\vector(0,1){8.00}}
\put(1.00,13.00){\makebox(0,0)[cc]{$t$}}
\put(14.00,0.00){\makebox(0,0)[cc]{$x$}}
\put(10.00,79.00){\makebox(0,0)[cc]{$t_A$}}
\put(91.00,32.00){\makebox(0,0)[cc]{$t_B$}}
\put(86.00,41.00){\circle{2.83}}
\put(86.00,26.00){\line(0,1){14.00}}
\put(74.00,20.00){\line(1,1){12.00}}
\put(15.00,79.00){\vector(3,-2){71.00}}
\end{picture}
\end{center}
\caption{Scheme of backward-in-time signalling by EPR-type telegraph.
The postulated controllability of outcomes in ${1}$, mediated
via ${2}$,
 is used
to transmit information. The flow of information is indicated by the
arrow. ``$\bullet$'' stands for the {\em active} mode; i.e.,
controllable outcome
(preparation). ``$\circ$'' stands for the {\em passive} mode; i.e.,
measurement. The two signs are drawn on top and at
bottom to indicate the orientation (relative angle $\pi$).
 \label{fig-0}}
\end{figure}
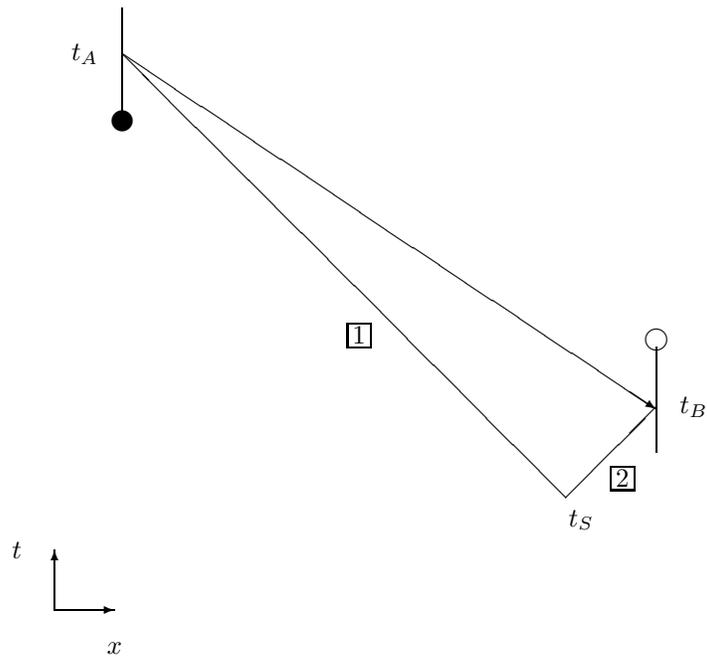

In what follows we shall consider the hypothetical transmission of
information backward in time. To be more specific, we shall use an
EPR-type telegraph which uses entangled particles
in
a singlet state (i.e.,
the total angular momentum of the two particles is zero)
as drawn in Fig.
\ref{fig-0}.
The apparatus is
tuned to convey perfect correlations of the direction of angular
momentum labelled by ``$+$'' and ``$-$''; i.e., the outcomes
are either $+\,+$ or $-\,-$. (Perfect correlations can be achieved by
choosing a relative angle of measurement of $\pi$.)
The (unphysical) assumption necessary for signalling backwards in
time is that on one side, say for particles in path ${1}$, the
{\em outcome can be controlled.}
This means that it will be assumed possible
to produce a particle with, say, direction of angular momentum
``$+$''
(``$-$'') in the path ${1}$ at $t_A$, thereby transmitting
a signal
``$+$'' (``$-$'') via its perfectly correlated entangled partner in
path ${2}$ to a second observer back in time at $t_B$; thereby,
$t_A>t_B>t_S$ but otherwise
arbitrary.

An alternative setup for backward in time signalling operates with
parameter dependence
\cite{shimony,shimony3}.
There, the (unphysical) assumption is that the measurement
outcomes in one
path depend on the setting of the measurement angle in the other path.

We shall make use of the EPR-type telegraph to construct a time paradox
and argue against parameter dependence
\cite{shimony,shimony3}
 and outcome
controllability
in any form.
In a similar
manner, the liar paradox \cite{bible} was translated by G\"odel into
arithmetic
\cite{godel} to argue against a complete description of a formal system
within that very system \cite{burks}.
For instance, the g\"odelian sentence \cite{popper-50}
claiming its own
unprovability in a particular system appears undecidable within that
very system.
In physical terms,
undecidability must be translated onto the level of {\em
phenomena.}
To put it pointedly: there is no such thing as an inconsistent
phenomenon.
In a {\em yes-no} experiment which can have two
possible outcomes, only one of these outcomes will actually be measured.
There might even be a ``hidden parameter  (extrinsic \cite{svozil:83},
exo-
\cite{roessler}) arena,'' in which this particular
outcome could be
deterministically accounted for. Yet,
for an intrinsic observer who is embedded in the system
\cite{toffoli:79},
this
level will be permanently inaccessible \cite{ro-cp}.
As will be shown below, quantum
mechanics implements this phenomenological undecidability both by the
postulate of randomness of certain outcomes and by the superposition
principle.
Related arguments have been put forward in
\cite{popper-50,rothstein-82,peres-84,wolfram1,moore,elitzur,posiewnik}.

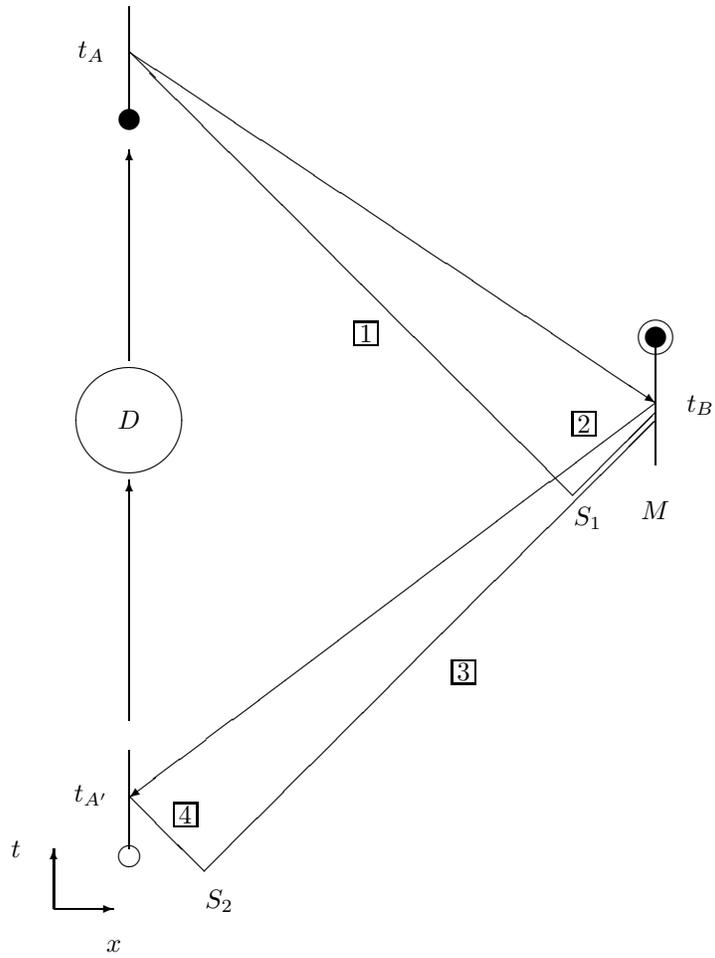
\begin{figure}
\begin{center}
\unitlength=1.00mm
\special{em:linewidth 0.4pt}
\linethickness{0.4pt}
\begin{picture}(92.00,125.00)
\put(16.00,20.00){\line(1,-1){10.00}}
\put(26.00,10.00){\line(1,1){60.00}}
\put(75.00,60.00){\line(-1,1){59.00}}
\put(77.00,57.00){\makebox(0,0)[cc]{$S_1$}}
\put(28.00,6.00){\makebox(0,0)[cc]{$S_2$}}
\put(75.00,60.00){\line(1,1){11.00}}
\put(16.00,119.00){\vector(3,-2){70.00}}
\put(86.00,72.33){\vector(-4,-3){70.00}}
\put(46.00,80.00){\framebox(3.00,3.00)[cc]{$1$}}
\put(86.00,58.00){\makebox(0,0)[cc]{$M$}}
\put(75.00,68.00){\framebox(3.00,3.00)[cc]{$2$}}
\put(59.00,35.00){\framebox(3.00,3.00)[cc]{$3$}}
\put(22.00,16.00){\framebox(3.00,3.00)[cc]{$4$}}
\put(16.00,125.00){\line(0,-1){15.00}}
\put(16.00,110.00){\circle*{2.83}}
\put(86.00,80.00){\line(0,-1){16.00}}
\put(16.00,12.00){\circle{2.83}}
\put(86.00,81.00){\circle{4.47}}
\put(86.00,81.00){\circle*{2.83}}
\put(16.00,13.00){\line(0,1){13.00}}
\put(16.00,70.00){\circle{14.00}}
\put(16.00,70.00){\makebox(0,0)[cc]{$D$}}
\put(16.00,30.00){\vector(0,1){32.00}}
\put(16.00,78.00){\vector(0,1){28.00}}
\put(6.00,5.00){\vector(1,0){8.00}}
\put(6.00,5.00){\vector(0,1){8.00}}
\put(1.00,13.00){\makebox(0,0)[cc]{$t$}}
\put(14.00,0.00){\makebox(0,0)[cc]{$x$}}
\put(11.00,119.00){\makebox(0,0)[cc]{$t_A$}}
\put(11.00,20.00){\makebox(0,0)[cc]{$t_{A'}$}}
\put(92.00,72.00){\makebox(0,0)[cc]{$t_B$}}
\end{picture}
\end{center}
\caption{time paradox.
Two backward-in-time signalling devices are used here, but only one
would be necessary, the other could be subluminal quantum information
channel. The important point is the outcome
controllability at $t_A$ with regards to the measurement at $t_{A'}$.
 \label{f-1}}
\end{figure}

Consider two
backward-in-time signalling
EPR-type telegraphs of the above type arranged
as drawn in Fig. \ref{f-1}.
Physically, the flow of information is mediated via the two entangled
pairs in paths {1}--{2} and {3}--{4}.
An information  in  2 is mirrored by $M$ in 3.
By this instrument, some mechanistic agent $A$ (e.g., computer,
deterministic observer) which is given the power of outcome control
or, alternatively, parameter dependence,
can exchange information
with itself on closed timelike lines
\cite{godel-rmp,recami-87,bell,deutsch-91}.
$A$ shall be confronted with the following paradoxical task.
Whenever $A$ registers the information ``$+$'' (``$-$'') at time
$t_{A'}$,
$A$ must stimulate the opposite outcome ``$-$'' (``$+$'') at the later
time
$t_A$.

Before discussing the paradox, let us consider the two states
$\vert 0\rangle \equiv$ ``$-$''
and
$\vert 1\rangle \equiv$ ``$+$'' which are accessible to $A$.
These states can be the basis of a cbit with the identification
of the
symbols ``$0$'' and ``$1$'' for
$\vert 0\rangle$
and
$\vert 1\rangle $,
 respectively.
Quantum mechanically any coherent superposition of them is allowed.
$A$'s paradoxical task can be formalized
by a unitary evolution operator
$
\widehat{D}
$ as follows
\begin{equation}
\widehat{D} \vert 0\rangle  = \vert 1\rangle , \quad
\widehat{D} \vert 1\rangle  = \vert 0\rangle \quad .
\end{equation}
In the state basis $\{ \vert 0\rangle , \vert 1\rangle \}$,
$\widehat{D}$ is just equivalent to the unary logical not-operation and
is therefore identical with the not-gate
(or the Pauli spin operator $\tau_1$),
\begin{equation}
\widehat{D}=
\tau_1 =
\left(
\begin{array}{cc}
0 & 1\\
1 & 0
\end{array}
\right) =\vert 1\rangle \langle 0\vert
+ \vert 0\rangle \langle 1\vert    \quad .
\end{equation}
The syntactic structure of the
paradox closely resembles Cantor's  diagonalization me\-thod
which has been
applied by G\"odel, Turing and others for undecidability proofs in a
recursion theoretic setup \cite{davis,rogers,odi:89,svozil-93}.
Therefore,
$
\widehat{D}
$
will be called
{\em diagonalization} operator,
despite the fact that its only nonvanishing components are
off-diagonal.
(Notice that $A$'s task would be perfectly consistent if
there were no ``bit switch'' and if thus
$
\widehat{D} =\mbox{ diag}(1,1)
$.)

The paradoxical feature of the construction reveals itself in the
following question: what happens to $A$?
In particular: what does $A$ register and send?

Let us first consider these questions from a classical perspective.
Classically, the particles with which $A$ operates can only be in one of
two possible
states, namely in $\vert 0\rangle $ or in $\vert 1\rangle $,
corresponding to the classical bit states.
By measuring the particle in beam 4, $A$ gets either the outcome ``$+$''
or ``$-$''.
In both cases, the agent $A$ is lead to a complete
contradiction.

For, if $A$
receives ``$+$'', corresponding to cbit state 1,
$A$ is obliged to send out ``$-$'', corresponding to cbit state 0
($A$ has been assumed to be able to control the outcomes in beam 1).
Due to the perfect EPR-correlations, the
partner particle in beam
2 is registered as
``$-$'' at the mirror at time $t_B$. By controlling the outcome in beam
3, this mirrored cbit can again be sent backwards in time, where ``$-$''
is received by
$A$ via a measurement of the particle in beam 4. This, however,
contradicts
the initial assumption that the outcome in beam 4 is ``$+$''.

On the other hand,
if $A$
receives ``$-$'', corresponding to cbit state 0,
 $A$ is obliged to send out ``$+$'', corresponding to cbit state 1;
yet, since at $t_B$ the cbit is just
reflected as described above,
$A$ should have received ``$+$''.
Thus classically, agent $A$ is in an inescapable dilemma.

The defense strategy in formal logic and classical recursion theory
against such inconsistencies is
to avoid the appearance of a paradox by claiming (stronger: requiring)
overall consistency, resulting in no-go theorems; i.e., in the
postulate of the impossibility of any operational method, procedure or
device which would have the potentiality to cause a paradox.
(Among the many impossible objects giving rise to paradoxes are
such seemingly innocent devices as a
``halting
 algorithm'' computing whether or not another
arbitrary computable algorithm produces a particular output;
or an algorithm identifying
another arbitrary algorithm by input-output experiments.)

In the above case, the defense strategy would result in the postulate of
the impossibility of any
backward-in-time
information flow or, more general,
of closed timelike lines.
Since the only nontrivial feature of the
backward-in-time
information flow
has been outcome controllability or parameter dependence,
the diagonalization
argument can be used against outcome controllability and parameter
dependence, resulting in an
intrinsic randomness of the individual outcomes.

Quantum mechanics implements exactly that kind of recursion theoretic
argument; yet in a form which is not common in recursion theory. Observe
that the paradox is resolved when
$A$ is allowed a nonclassical qbit of information.
In particular, $A$'s task can consistently be performed
if it inputs a qbit corresponding to the {\em fixed point} state
of
$\widehat{D}$; i.e.,
\begin{equation}
\widehat{D}\vert \ast \rangle =\vert \ast \rangle \quad .
\end{equation}
The fixed point state $\vert \ast \rangle $ is
just the eigenstate of the diagonalization operator
$\widehat{D}$ with
eigenvalue $1$.
Notice that the eigenstates of
$\widehat{D}$ are
\begin{equation}
\vert I\rangle      ,
\vert II\rangle
=
{1\over \sqrt{2}}\left[
\left( \begin{array}{c}1\\ 0\end{array} \right)\pm
\left( \begin{array}{c}0\\ 1\end{array} \right)
\right]
=
{1\over \sqrt{2}}(\vert 0\rangle \pm \vert 1\rangle )
\end{equation}
with the  eigenvalues $+1$ and $-1$, respectively.
Thus, the nonparadoxical, fixed point qbit
in the basis of $\vert 0\rangle $ and $\vert 1\rangle $ is given by
\begin{equation}
\vert \ast \rangle =\vert {1\over \sqrt{2}},{1\over \sqrt{2}} \rangle
=\vert I\rangle \quad .
\end{equation}
In natural language,
this qbit solution corresponds to the statement that
it is impossible
for the agent to control the outcome, since
 there is a
fifty percent chance for the  classical bit states $\vert 0\rangle$ and
$\vert 1\rangle$ to be ``stimulated'' at
$t_{A}$.
The impossibility of outcome control
(and parameter independence) is
indeed encountered
in quantum mechanics \cite{fp}.

We close the discussion on the consistent use of paradoxes in physics
with a few comments. First, it is important to recognize that the above
considerations have {\em no} immediate bearing on quantum
complementarity.
In the author's opinion, complementarity is a general feature of the
intrinsic perception of computer-generated universes, which is
realizable already at a very elementary pre-diagonalization level
\cite{e-f-moore,finkelstein-83,svozil-93}; i.e., without
the requirement of computational universality or its arithmetic
equivalent.

As has been pointed out before,
the above argument remains valid
for any conceivable (local or nonlocal
\cite{pitowsky-82,pitowsky})
 hidden
variable theory.
The consistency of the physical
phenomenology requires that hidden variables remain inaccessible to an
intrinsic observer.
Pointedly stated,
from an intrinsic, operational point of view, when re-interpreted
properly, a paradox
marks the appearance of uncertainty and uncontrollability (cf. a
statement by G\"odel \cite{burks}).

A similar treatment of the halting problem \cite{rogers} for a quantum
computer leads to the conclusion that the quantum recursion theoretic
``solution'' of the
halting problem
reduces to the tossing of a fair (quantum \cite{svozil111}) coin
\cite{chaitin}.
Another, less abstract, application for quantum information theory is
the handling of inconsistent information in databases.
Thereby,
two contradicting cbits of information
$\vert a\rangle $ and
$\vert b\rangle $ are resolved by the qbit
$\vert {1/ \sqrt{2}},{1/ \sqrt{2}} \rangle =
({1/ \sqrt{2}})(\vert a\rangle + \vert b\rangle )$.
Throughout the rest of the computation the coherence is maintained
\cite{fuzzy}. After the processing, the result is obtained by a
measurement. The processing of qbits requires an exponential space
overhead on classical computers in cbit base \cite{feynman}.
Thus, in order to remain tractable,
the corresponding qbits should be implemented on
truly quantum universal computers.

Acknowledgements:
This and related topics were vigorously discussed in seminars and
regular
Viennese coffee house sessions with Anton Zeilinger, Johann Sumhammer
and others. G\"unther Krenn took the pain to read, contribute and
comment to several versions of the manuscript. An anonymous referee
suggested many revisions.
Nonetheless,
any blame should remain solely with the author.

\end{document}